# Chiral sensing with achiral isotropic metasurfaces

Sotiris Droulias

*Institute of Electronic Structure and Laser, FORTH, 71110 Heraklion, Crete, Greece*

Metasurfaces, the two-dimensional analogues of metamaterials, are ideal platforms for sensing molecular chirality at the nanoscale, e.g. of inclusions of natural optically active molecules, as they offer large accessible areas (they are essentially surfaces) and can accommodate the necessary strong resonances for coupling the probing radiation with the chiral inclusions. Here, we examine theoretically achiral isotropic metasurfaces, and we treat them as polarizable surfaces that support resonant electric and magnetic currents, which are coupled via the chiral inclusions. We derive analytically, and verify numerically, expressions that provide insight to the enhancement mechanism of the magneto-electric coupling and explain why circular dichroism signals (difference in absorption between left- and right- circularly polarized waves) can arise from both the real and the imaginary part of the chirality parameter $\kappa$. Our analysis demonstrates distinct chiroptical signals where the contributions from both the real and imaginary part of $\kappa$ can be independently observed and, based on such measurements, we propose a scheme for the unambiguous determination of an unknown chirality.

## I. INTRODUCTION

Chirality, a geometric property in which an object is non-superimposable with its mirror image, is frequently met in many biomolecules and chemical compounds, which can exist in right- and left-handed forms, known as enantiomers. Their functionality is often determined by their handedness and, therefore, the ability to efficiently sense molecular chirality is of fundamental importance for many research disciplines and industries, such as the agricultural, pharmaceutical, and chemical [1].

A chiral medium is described electromagnetically by the constitutive relations $\mathbf{D}=\varepsilon_0\varepsilon\mathbf{E}-i(\kappa/c)\mathbf{H}$, $\mathbf{B}=\mu_0\mu\mathbf{H}+i(\kappa/c)\mathbf{E}$ [2], where $\varepsilon,\mu$ are the relative permittivity and permeability ($\varepsilon_0,\mu_0$ are the vacuum permittivity and permeability) and $c$ the vacuum speed of light; $\kappa$ is the chirality (also known as 'Pasteur') parameter, which expresses the chiral molecular response. The magnitude of the chirality parameter, $|\kappa|$, is a function of the molecular properties (i.e. polarizability) and its concentration, while its sign, $\text{sgn}(\kappa)$, depends on the handedness of the medium [3]. Therefore, in order to be able to detect and distinguish enantiomers and quantify their chiral response, a chiral sensing scheme should be sensitive to $\text{sgn}(\kappa)$ and $|\kappa|$, respectively.

Based on the fact that the onset of chirality ($\kappa\neq 0$) induces magneto-electric coupling and, therefore, a nonvanishing pseudoscalar $\mathbf{D}\cdot\mathbf{B}$ product, contemporary nanophotonic approaches for chiral sensing aim to enhance the chiral wave-matter interaction by preparing an environment of strong local fields with parallel electric and magnetic components (and proper phase lag). Usually this is quantified with $C$, the optical chirality density, which in its time-averaged form is given by $C = -(\omega/2)\text{Im}(\mathbf{D}^*\mathbf{B})$ [4,5]; here $\omega$ is the angular frequency and the asterisk denotes the complex conjugate. Following this prescription, several approaches have been proposed to provide strong interaction between the chiral molecules to be probed (we will call them *'chiral inclusions'*, in general) and the nanophotonic system, such as propagating surface plasmons [6,7], plasmonic particles [8–16], chiral metamaterials [17–23] and, recently, achiral metamaterials [24–32].

Intuitively, chiral metamaterials seem to be the most promising candidate, as they are based -by principle of operation- on the excitation of strong parallel, and appropriately phased, **D** and **B** components, thus providing near fields with extremely enhanced $C$ (compared to that provided by far-field circularly polarized plane waves). However, $C$ changes sign across each unit cell [17], thus resulting in a reduced average value. If the chiral substance can be placed at certain hot spots (similarly to nonchiral configurations of plasmonic [15] and dielectric [33] nanoparticles), the enhancement can indeed be significant. For practical applications, though, it would be beneficial to have a uniform and easily accessible sensing area.

Recently, it was shown that the chirality of the metasurface itself is not necessary, and achiral metasurfaces can circumvent the problem of reduced averaged $C$ [25,27,29,32]. The principle of operation is based on appropriately tailoring two modes of the metasurface that provide strong –collinear– dipole moments, one of electric type and one of magnetic type (we will call them *'electric mode'* and *'magnetic mode'*). Illumination with circularly polarized plane waves excites both modes simultaneously, subsequently producing the necessary strong $\mathbf{D}\cdot\mathbf{B}$ product, for probing the chiral inclusions. While this approach explains the chiral sensing mechanism on the basis that $\kappa$ couples to both modes simultaneously, it would be instructive to consider an alternative perspective: the two modes are coupled *via* $\kappa$, as demonstrated in [31] under illumination with linearly polarized waves. As both approaches are equivalent -the latter having been extensively overlooked thus far-, there is a need for a unified theoretical description that can reveal both the potential and limits of chiral sensing with achiral metasurfaces, especially in view of the large volume of relevant numerical and experimental works [24–32].

In this work we give a simple theoretical model for the description of achiral isotropic metasurfaces with chiral inclusions that elucidates how aspects of chiral sensing are associated with the properties of the metasurfaces. The model is based on replacing the actual metasurface with a polarizable sheet that supports electric and magnetic surface currents. Since the metasurface has no inherent chirality, the magneto-electric coupling comes entirely from the coupling of the two modes via the chiral inclusions. We show analytically, and verify numerically, that the magneto-electric coupling of the composite system, i.e. metasurface with chiral inclusions, can be expressed as a product of three quantities, namely (a) the chirality parameter $\kappa$ of the chiral inclusions, (b) the spatial overlap of the electric and magnetic mode of the metasurface and (c) the individual response functions (surface conductivities) of the two modes. As a direct consequence, we show that the far-field measurements are proportional to the chirality parameter $\kappa$ of the chiral inclusions, thus justifying why the identification of both sgn($\kappa$) and |$\kappa$| is possible. Additionally, we show that, because the magneto-electric coupling of the composite system involves contributions from both the chiral inclusions and the metasurface, a difference in the absorption between left- and right-circularly polarized waves (i.e. circular dichroism –or CD) can result from both the real and the imaginary part of $\kappa$, Re($\kappa$) and Im($\kappa$), respectively. This is a surprising result, because in several contemporary approaches CD signals are predicted to be sensitive to Im($\kappa$) only [7,15–17,20,24–30]; however, our finding is in accord with observations in recent experiments [14,32], and explains the origin of enhanced CD signals especially at frequencies far from the chiral molecular resonances, where Im($\kappa$) is weak and Re($\kappa$) is dominant. Last, we explain how it is possible to detect and distinguish Re($\kappa$) and Im($\kappa$) with measurements of the optical rotation and ellipticity of the transmitted wave and we propose a scheme for the unambiguous determination of an unknown chirality, based on such measurements.

## II. THEORETICAL FORMULATION

### A. The coupled oscillator model

Let us consider an achiral isotropic metasurface with a homogeneously embedded chiral inclusion; the metasurface supports one electric and one magnetic mode and the presence of the chiral inclusion induces weak coupling between the two –otherwise orthogonal– modes. In order to understand the coupling mechanism, we can replace the metasurface with a polarizable sheet (zero thickness) that supports two collinear current densities, one of electric type, $j_e$, and one of magnetic type, $j_m$ (see Fig. 1). Then we can write a coupled oscillator model for $j_e$ and $j_m$ as described in [31] (there the equivalent model was based on electric and magnetic dipole moments $p$ and $m$, which are directly related to $j_e$ and $j_m$ as $j_e = i\omega p$ and $j_m = i\omega m$). This approach allows for concise analytical expressions, as the quantities involved are surface quantities, i.e. the spatial dimension is eliminated. To relate the microscopic quantities described in the coupled oscillator model with the macroscopic fields we need to express the electric and magnetic surface (complex) conductivities of the macroscopic medium in terms of the local (surface) current densities $j_e$, $j_m$, and the local (surface) fields $E_{loc}$, $H_{loc}$ (at the location of the polarizable sheet). In matrix form this reads:

$$\begin{pmatrix} \langle j_e \rangle \\ \langle j_m \rangle \end{pmatrix} = \begin{pmatrix} \sigma_{ee} & \sigma_{em} \\ \sigma_{me} & \sigma_{mm} \end{pmatrix} \begin{pmatrix} E_{loc} \\ H_{loc} \end{pmatrix} \qquad (1)$$

where the brackets denote spatial average and $\sigma_{ee}$, $\sigma_{mm}$, $\sigma_{me}$, $\sigma_{em}$ are the surface conductivities to be retrieved from the coupled oscillator model. In each $\sigma_{ij}$ the subscript $\{i,j\} = \{e,m\}$ denotes the electric 'e' or magnetic 'm' character of the associated conductivities and fields (subscript 'i' and 'j', respectively). The averaged surface electric and magnetic currents, $\langle j_e \rangle$ and $\langle j_m \rangle$, are measured in A/m and V/m, respectively; the electric and magnetic conductivities, $\sigma_{ee}$ and $\sigma_{mm}$, are measured in S and $\Omega$, respectively, and $\sigma_{em}$, $\sigma_{me}$ are dimensionless [see Supporting Information (SI) for details]. The conductivities $\sigma_{em}$, $\sigma_{me}$ express the magneto-electric coupling induced by the chiral inclusions, and are zero in the absence of chirality. Because in natural chiral media the bulk chirality parameter $\kappa$ is ~4-6 orders of magnitude weaker than the susceptibilities of typical dielectrics and metamaterials [14,34,35], $\kappa$ is practically a perturbation to the strong resonant response of the metasurface. Therefore, as we will show, the onset of chirality leaves $\sigma_{ee}$, $\sigma_{mm}$ practically unchanged and the chiral contributions can be introduced in $\sigma_{em}$, $\sigma_{me}$ entirely. This allows us first to determine $\sigma_{ee}$, $\sigma_{mm}$ for the metasurface without chiral inclusions and subsequently use $\sigma_{ee}$, $\sigma_{mm}$ to find $\sigma_{em}$, $\sigma_{me}$ once the chiral substance is introduced.

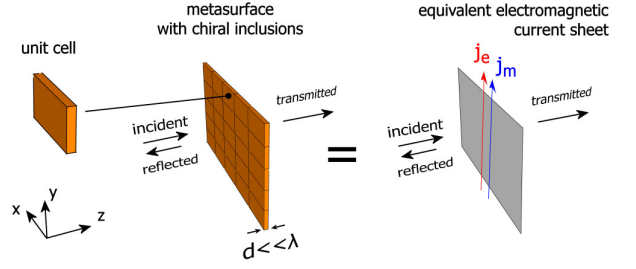

Figure 1: Schematic of achiral metasurface of thickness $d \ll \lambda$ ($\lambda$: wavelength) for enhanced sensing of chiral inclusions. The metasurface supports modes with electric and magnetic moments that interact with the chiral inclusions. These moments are equivalently described by an infinite electromagnetic current sheet supporting an electric current $\mathbf{j_e}$ and a magnetic current $\mathbf{j_m}$, which are coupled via a magneto-electric conductivity.

Let us start with the metasurface without chiral inclusions; this corresponds to a set of two uncoupled equations, one for each current density:

$$\frac{d^2 j_e(t)}{dt^2} + \gamma_e \frac{d j_e(t)}{dt} + \omega_e^2 j_e(t) = f_e(t)$$
$$\frac{d^2 j_m(t)}{dt^2} + \gamma_m \frac{d j_m(t)}{dt} + \omega_m^2 j_m(t) = f_m(t) \quad (2)$$

where $\omega_e$, $\omega_m$ are the resonant frequencies, $\gamma_e$, $\gamma_m$ the damping rates and $f_e$, $f_m$ the driving forces of $j_e$ and $j_m$, respectively. Assuming a solution of the form $j_e(t) = \tilde{j}_e(\omega)\exp(i\omega t), j_m(t) = \tilde{j}_m(\omega)\exp(i\omega t)$ Eqs. (2) can be solved in the frequency domain:

$$\tilde{j}_e(\omega) = \frac{1}{D_e(\omega)} \tilde{f}_e(\omega)$$
$$\tilde{j}_m(\omega) = \frac{1}{D_m(\omega)} \tilde{f}_m(\omega) \quad (3)$$

where $D_q(\omega) = \omega_q^2 - \omega^2 + i\gamma_q\omega$, $q = \{e,m\}$. We can now use these expressions to find the surface conductivities $\sigma_{ee}$, $\sigma_{mm}$, by relating the microscopic quantities $\tilde{j}_e, \tilde{j}_m, \tilde{f}_e, \tilde{f}_m$ to the macroscopic quantities $\langle \tilde{j}_e \rangle, \langle \tilde{j}_m \rangle, \tilde{E}_{loc}, \tilde{H}_{loc}$; in the absence of magneto-electric coupling, the latter are related as $\langle \tilde{j}_e \rangle = \sigma_{ee}\tilde{E}_{loc}$ and $\langle \tilde{j}_m \rangle = \sigma_{mm}\tilde{H}_{loc}$ [see Eq.(1)]. Following the procedure analyzed in [36], we first note that if there are $n$ atoms per unit of surface area, the average surface current density equals $\langle \tilde{j}_e \rangle = n\tilde{j}_e$ and $\langle \tilde{j}_m \rangle = n\tilde{j}_m$, respectively, since $j_e$, $j_m$ is the contribution of each constituent meta-atom to the total current density of the metasurface. Additionally, because the (microscopic) driving forces $f_e$, $f_m$ drive currents rather than polarizations, they are proportional to the derivatives of the (macroscopic) fields $E_{loc}$ and $H_{loc}$, respectively, rather than the fields themselves, that is, $\tilde{f}_e = i\omega C_e \tilde{E}_{loc}$ and $\tilde{f}_m = i\omega C_m \tilde{H}_{loc}$. To find the proportionality constants $C_e$, $C_m$ we can recall that, for our linear meta-atom, each average surface current must be proportional to the corresponding field at the surface, that is: $\langle \tilde{j}_e \rangle = \sigma_{ee}\tilde{E}_{loc}$ or $n\tilde{j}_e = \sigma_{ee}\tilde{E}_{loc}$ or $nD_e^{-1}\tilde{f}_e = i\omega\varepsilon_0\chi_{ee}\tilde{E}_{loc}$ where we have used the result of Eq. (3) and the relation $\sigma_{ee} = i\omega\varepsilon_0\chi_{ee}$ which connects the surface electric conductivity with the respective surface susceptibility. Comparing the result $\tilde{f}_e = n^{-1}D_e i\omega\varepsilon_0\chi_{ee}\tilde{E}_{loc}$ with $\tilde{f}_e = i\omega C_e \tilde{E}_{loc}$, we obtain in the static limit ($\omega \to 0$): $C_e = n^{-1}\omega_e^2\varepsilon_0\chi_{ee}^{(static)}$, as has also been retrieved previously in [36]. Following a similar procedure for $C_m$ we find $C_m = n^{-1}\omega_m^2\mu_0\chi_{mm}^{(static)}$ and we can now write:

$$\sigma_{ee} = \frac{\langle \tilde{j}_e \rangle}{\tilde{E}_{loc}} = \frac{n\tilde{j}_e}{(i\omega C_e)^{-1}\tilde{f}_e} = \frac{i\omega a_e}{D_e(\omega)}$$
$$\sigma_{mm} = \frac{\langle \tilde{j}_m \rangle}{\tilde{H}_{loc}} = \frac{n\tilde{j}_m}{(i\omega C_m)^{-1}\tilde{f}_m} = \frac{i\omega a_m}{D_m(\omega)} \quad (4)$$

where $a_e \equiv nC_e = \omega_e^2\varepsilon_0\chi_{ee}^{(static)}$, $a_m \equiv nC_m = \omega_m^2\mu_0\chi_{mm}^{(static)}$.

Next we introduce a weak coupling coefficient $\kappa_c$, to account for the magneto-electric coupling between $j_e$ and $j_m$ and the system of Eq.(2) takes the form:

$$\frac{d^2 j_e(t)}{dt^2} + \gamma_e \frac{dj_e(t)}{dt} + \omega_e^2 j_e(t) - \frac{\kappa_c}{nC_m} j_m(t) = f_e(t)$$
$$\frac{d^2 j_m(t)}{dt^2} + \gamma_m \frac{dj_m(t)}{dt} + \omega_m^2 j_m(t) + \frac{\kappa_c}{nC_e} j_e(t) = f_m(t) \quad (5)$$

Because $\kappa_c$ connects electric and magnetic quantities, rather than only electric quantities as in [36], the normalization in $\kappa_c$ is chosen to preserve the correct units among the involved quantities, that is: $j_e$ [A×m], $j_m$ [V×m], $n$ [1/m²], $C_e$ [S×Hz×m²], $C_m$ [Ω×Hz×m²], and $\kappa_c$ [Hz³]. The ± sign in front of $\kappa_c$ reflects its connection with $\kappa$ ($j_e$ and $j_m$ are coupled via $\kappa_c$ similarly to how **E** and **D** are coupled to **B** and **H**, respectively, via $\kappa$). Equations (5) can be solved in the frequency domain to yield:

$$\tilde{j}_e(\omega) = \overbrace{\frac{D_m}{D_e D_m + \kappa_c^2 a_e^{-1} a_m^{-1}}}^{\tilde{j}_{ee}} \tilde{f}_e + \overbrace{\frac{\kappa_c a_m^{-1}}{D_e D_m + \kappa_c^2 a_e^{-1} a_m^{-1}}}^{\tilde{j}_{em}} \tilde{f}_m$$
$$\tilde{j}_m(\omega) = \underbrace{\frac{D_e}{D_e D_m + \kappa_c^2 a_e^{-1} a_m^{-1}}}_{\tilde{j}_{mm}} \tilde{f}_m - \underbrace{\frac{\kappa_c a_e^{-1}}{D_e D_m + \kappa_c^2 a_e^{-1} a_m^{-1}}}_{\tilde{j}_{me}} \tilde{f}_e \quad (6)$$

where we have used $a_e \equiv nC_e$, $a_m \equiv nC_m$, as introduced in Eq.(4). Due to the extremely weak magneto-electric coupling we can approximate $D_e D_m + \kappa_c^2 a_e^{-1} a_m^{-1} \cong D_e D_m$. Under this approximation, $\tilde{j}_{ee}, \tilde{j}_{mm}$ become identical to $\tilde{j}_e, \tilde{j}_m$ of the uncoupled case [Eq.(3)] and $\sigma_{ee}$, $\sigma_{mm}$, are therefore given by Eq.(4). We emphasize that for stronger coupling (as in chiral metamaterials for example) this approximation does not hold. For the remaining $\sigma_{em}$, $\sigma_{me}$ we may write:

$$\sigma_{em} = \frac{\langle \tilde{j}_{em} \rangle}{\tilde{H}_{loc}} = \frac{n\tilde{j}_{em}}{(i\omega C_m)^{-1}\tilde{f}_m} = +\frac{i\omega\kappa_c}{D_e(\omega)D_m(\omega)}$$
$$\sigma_{me} = \frac{\langle \tilde{j}_{me} \rangle}{\tilde{E}_{loc}} = \frac{n\tilde{j}_{me}}{(i\omega C_e)^{-1}\tilde{f}_e} = -\frac{i\omega\kappa_c}{D_e(\omega)D_m(\omega)} \quad (7)$$

thereby concluding the derivation of the macroscopic surface conductivities in terms of the microscopic quantities of the coupled oscillator model.

## B. The magneto-electric conductivity $\sigma_c$

The result of Eq.(7) clearly shows that the magneto-electric coupling induced by the chiral inclusions can be enhanced by the metasurface's resonant modes. To gain further insight into the enhancement mechanism, we combine Eq.(7) with Eq.(4), to express $\sigma_{em}$, $\sigma_{me}$ in terms of $\sigma_{ee}$, $\sigma_{mm}$:

$$\sigma_c \equiv \sigma_{em} = -\sigma_{me} = \frac{i\omega\kappa_c}{(i\omega a_e)(i\omega a_m)}\sigma_{ee}\sigma_{mm} \quad (8)$$

According to this result, in the weak-coupling regime, the magneto-electric conductivity $\sigma_c$ is proportional to $\sigma_{ee}$, $\sigma_{mm}$. In Eq.(8), all quantities are macroscopic, except for $\kappa_c$, which is the coupling constant in the coupled oscillator model. To express $\kappa_c$ in terms of macroscopic quantities as well, we need to find a connection with the coupling of the macroscopic fields, which is induced by the chiral inclusion.

Let us denote with ($\mathbf{E}_e$, $\mathbf{H}_e$) and ($\mathbf{E}_m$, $\mathbf{H}_m$) the fields of the electric and magnetic mode, respectively, that drive the local surface currents $j_e$ and $j_m$. In the absence of chirality ($\kappa=0$) the two modes are orthogonal and $j_e$, $j_m$ are therefore uncoupled. The onset of chirality ($\kappa\neq 0$) induces perturbative polarization densities of electric type $\delta\mathbf{P} = -i(\kappa/c)\mathbf{H}$, and of magnetic type $\delta\mathbf{M} = +i(\kappa/c)\mathbf{E}$, which cause the two modes to become coupled (and similarly $j_e$, $j_m$). The power transferred from mode $e$ to mode $m$ is caused by the polarization currents $i\omega\delta\mathbf{P}_{me} = +(\kappa\omega/c)\mathbf{H}_e$ and $i\omega\delta\mathbf{M}_{me} = -(\kappa\omega/c)\mathbf{E}_e$ and hence the total power transferred to mode $m$ can be found by integrating the quantity $i\omega\delta\mathbf{P}_{me}\ \mathbf{E}_m^* + i\omega\delta\mathbf{M}_{me}\ \mathbf{H}_m^*$, across the whole volume $V_{inclusion}$ where the chiral substance extends [37] (the total power transferred from mode $m$ to mode $e$ is found similarly and corresponds to exchanging $e\leftrightarrow m$ in the subscripts). Then, because the surface current densities are proportional to the local fields, the coupling coefficient $\kappa_c$ will be proportional to the integral:

$$\kappa_c \propto \int_{V_{inclusion}} \kappa\frac{\omega}{c}\left(\mathbf{E}_m^*\mathbf{H}_e - \mathbf{E}_e\mathbf{H}_m^*\right)dV \quad (9)$$

According to this result, which has also been retrieved previously in the framework of coupled-mode theory [38], the coupling coefficient is proportional to the chirality parameter $\kappa$ and an overlap integral of the fields, i.e. $\kappa_c$ is decomposed into the individual contributions from the chiral inclusion and the metasurface. Hence, the magneto-electric coupling of Eq.(8) can be simply expressed as:

$$\sigma_c \equiv \kappa C_0 \sigma_{ee}\sigma_{mm} \quad (10)$$

where $C_0$ is a constant containing the overlap integral of the fields and we have assumed relatively sharp resonances so that the angular frequency $\omega$ appearing explicitly in Eqs.(8),(9) is practically constant. This is one of the central results in this work; it captures the coupling between the two modes, as expressed via their response functions (surface conductivities). In particular, this result shows that the magneto-electric coupling of the composite metasurface-chiral inclusion is proportional to (a) the chirality parameter $\kappa$ of the chiral inclusion, (b) the spatial overlap of the electric and magnetic mode, and (c) the product of their individual conductivities, $\sigma_{ee}\sigma_{mm}$. In essence, $\sigma_c$ corresponds to the *effective chirality* of the composite metasurface-chiral inclusion (see SI), which is found here to be the chirality $\kappa$ of the chiral inclusion, multiplied by a factor $C_0\sigma_{ee}\sigma_{mm}$. As $C_0$, $\sigma_{ee}$ and $\sigma_{mm}$ are properties of the metasurface, they can be tailored independently from the chiral inclusion to enhance the product $C_0\sigma_{ee}\sigma_{mm}$, and, in turn, the weak chirality parameter $\kappa$. This can be achieved, for example, by choosing to work with electric and magnetic modes that overlap spatially in the region of the chiral inclusions and by tuning their individual frequencies to coincide.

## III. NUMERICAL EXPERIMENTS

### A. Retrieval of conductivities

Having established the connection between our microscopic model and the macroscopic conductivities, we need now to relate the latter with the response of realistic achiral isotropic metasurfaces with chiral inclusions. As $\sigma_c$ is dimensionless by definition, it is also convenient to introduce the dimensionless conductivities $s_{ee} = \zeta\sigma_{ee}/2$ and $s_{mm} = \sigma_{mm}/2\zeta$, where $\zeta = (\mu_0\mu/\varepsilon_0\varepsilon)^{1/2}$ is the wave impedance of the surroundings, where, without loss of generality, we have assumed a uniform surrounding environment. Then, the conductivity tensor of Eq.(1) takes the form:

$$\hat{\sigma} = \begin{pmatrix} \sigma_{ee} & \sigma_{em} \\ \sigma_{me} & \sigma_{mm} \end{pmatrix} \equiv \begin{pmatrix} \frac{2}{\zeta}s_{ee} & \sigma_c \\ -\sigma_c & 2\zeta s_{mm} \end{pmatrix} \quad (11)$$

To investigate how the chiral inclusion mediates the coupling between the two modes, we consider illumination with linearly polarized incident waves, rather than circularly polarized waves, as discussed in [31]. Using the conductivity tensor of Eq.(11) we solve Maxwell's equations with the appropriate boundary conditions, which are formulated as $\mathbf{n}\times(\mathbf{E_2} - \mathbf{E_1}) = -\langle\mathbf{j_m}\rangle$, $\mathbf{n}\times(\mathbf{H_2} - \mathbf{H_1}) = +\langle\mathbf{j_e}\rangle$, where $\mathbf{n}$ is the surface normal of the current sheet pointing from region 1 to region 2. We find that the transmission and reflection amplitudes are expressed in terms of the surface conductivities as:

$$t \equiv t_{xx} = t_{yy} = \frac{1 - s_{ee}s_{mm}}{(1 + s_{ee})(1 + s_{mm})} \quad (12a)$$

$$r \equiv r_{xx} = r_{yy} = \frac{s_{mm} - s_{ee}}{(1 + s_{ee})(1 + s_{mm})} \quad (12b)$$

$$t_c \equiv t_{xy} = -t_{yx} = \frac{\sigma_c}{(1 + s_{ee})(1 + s_{mm})} \quad (12c)$$

$$r_c \equiv r_{xy} = r_{yx} = 0 \quad (12d)$$

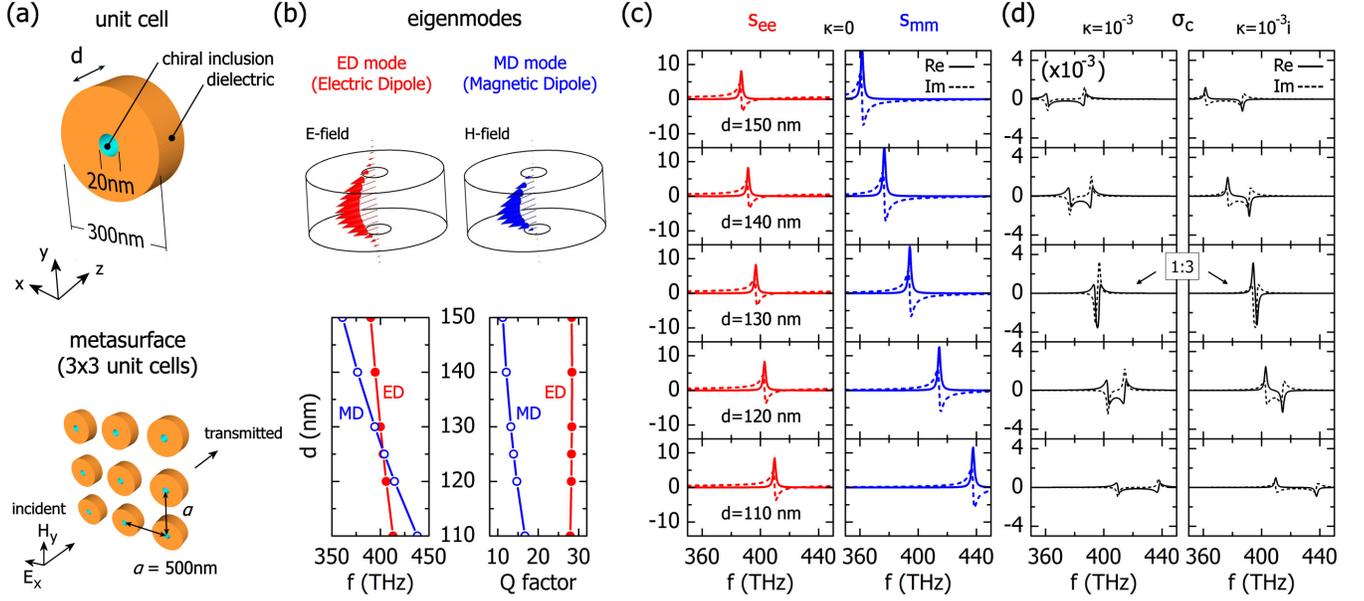

Figure 2: Achiral isotropic metasurface for enhanced chiral sensing. (a) Schematic of a single unit cell (top) and of the metasurface (bottom). (b) *Top panel*: field distribution of the electric-type (ED) and magnetic-type (MD) modes along the axis of the disk, where the chiral inclusion is located. *Bottom panel*: spectral tuning (left panel) and $Q$ factors (right panel) of ED and MD as a function of the disk height, $d$. (c) Retrieved conductivities $s_{ee}$, $s_{mm}$ as a function of the disk height, $d$, for $\kappa = 0$. (d) Retrieved conductivity $\sigma_c$ as a function of the disk height, $d$, for $\kappa = 10^{-3}$ (left panel) and $\kappa = 10^{-3}i$ (right panel). The results of $\sigma_c$ for $d = 130$ nm have been divided by a factor of 3 for easier comparison.

or, solving in terms of the conductivities:

$$s_{ee} = \frac{\zeta \sigma_{ee}}{2} = \frac{1-r-t}{1+r+t} \quad (13a)$$

$$s_{mm} = \frac{\sigma_{mm}}{2\zeta} = \frac{1+r-t}{1-r+t} \quad (13b)$$

$$\sigma_c = \frac{4t_c}{(1-r+t)(1+r+t)} \quad (13c)$$

where the subscripts in $t_{out,inc}$, $r_{out,inc}$ denote the output and incident polarization, respectively and we will call $r,t$ and $r_c,t_c$ the co- and cross- reflection/transmission amplitudes, respectively (for the general case where the refractive index of the surroundings changes across the metasurface see SI for the full analytical expressions). To derive Eqs.(12a)-(12d) and Eqs.(13a)-(13c) we made the crucial approximation $\sigma_c \ll s_{ee}, s_{mm}$, thus eliminating any $\sigma_c^2$ term. This approximation is valid, as long as the magneto-electric coupling is perturbative, as is the case here. The results of Eqs.(12a),(12b),(13a),(13b) are in accord with those previously retrieved in [39] and [40] in the context of achiral electromagnetic current sheets with isotropic electric/magnetic conductivities. The additional Eqs.(12c),(12d),(13c) account for the effect of magneto-electric coupling considered here.

To derive our analytical results we assumed that a realistic metasurface with chiral inclusions can be replaced by an equivalent electromagnetic sheet that has the same response, and can be described entirely in terms of surface conductivities. In other words, both the metasurface with the chiral inclusions and the electromagnetic sheet are described by the same reflection and transmission amplitudes $r$, $r_c$, $t$, $t_c$ as expressed by Eqs.(12a)-(12d). Therefore, we can use $r$, $r_c$, $t$, $t_c$ from the simulations of the metasurface to retrieve the surface conductivities $s_{ee}$, $s_{mm}$, $\sigma_c$ of its equivalent sheet model with Eqs.(13a)-(13c); then we can use the latter to understand the mechanisms of chiral sensing in the realistic system to which it corresponds and to retrieve the unknown chirality parameter, $\kappa$, of the chiral inclusions. Importantly, because the chirality parameter $\kappa$ of natural optically active media is dispersive, $\sigma_c$ exhibits the dispersive features of both the chiral inclusions and the metasurface [see Eq.(10)]. Therefore, in order to isolate the contribution from the metasurface we will start with a constant $\kappa$, which is a reasonable approximation when studying metasurfaces with resonances of much narrower linewidth than that of the chiral inclusions.

As an example, let us consider an achiral metasurface composed of a two-dimensional periodic arrangement of silicon nanodisks (refractive index $3.5-0.01i$) as the one examined in [25] [see Fig. 2(a)]. The nanodisks have radius 150nm, they are periodically arranged with a lattice constant of 500 nm and they have cylindrical holes of 10nm radius, in which the chiral inclusion is contained (refractive index $1.33-0.001i$). The system, for simplicity, is embedded in air. The disk height $d$ is used for tuning the spectral separation between the two modes, which we label as ED (electric dipole mode) and MD (magnetic dipole mode). Their field distribution along the center of the nanodisk is shown in Fig. 2(b), top panel, where also their frequency tuning and individual $Q$ factors are also shown [Fig. 2(b), bottom]. This is the outcome of full-wave vectorial Finite Element Method (FEM) simulations, performed with the commercial software COMSOL Multiphysics. To examine the response of the metasurface, first, with $\kappa = 0$ we illuminate the system with linearly polarized light as shown in Fig. 2(a) and calculate the reflection and transmission amplitudes for disk heights ranging from 110 nm to 150 nm. Using these results we retrieve the surface conductivities $s_{ee}$, $s_{mm}$ [with Eqs.(13a),(13b)], which we plot in Fig. 2(c). Next, we switch on $\kappa$ and repeat the calculations to retrieve $\sigma_c$ [using Eq.(13c)]. The results are shown in Fig. 2(d), left panel for $\kappa = 10^{-3}$ and in Fig. 2(d), right panel for $\kappa = 10^{-3}i$. We observe that $\sigma_c$ is maximized when the detuning between $s_{ee}$ and $s_{mm}$ is minimized, which occurs for $d = 130$nm. Additionally, we observe that the change from purely real to purely imaginary $\kappa$ results in $\sigma_c(\omega; \kappa = 10^{-3}i) = i\sigma_c(\omega; \kappa = 10^{-3})$.

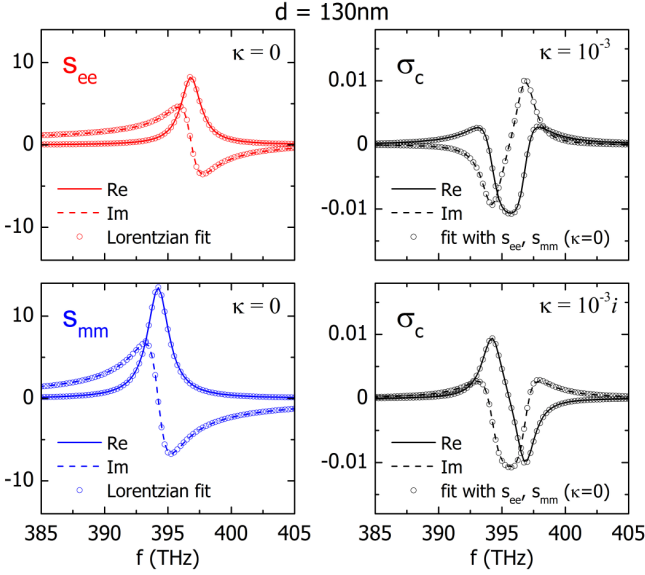

Figure 3: Numerically retrieved conductivities for the achiral isotropic metasurface of Fig. 2 with $d = 130$ nm. *Left column*: electric $s_{ee}$ and magnetic $s_{mm}$ conductivity for $\kappa = 0$ and fit with Lorentzian function (open circles). *Right column*: magneto-electric conductivity $\sigma_c$ for $\kappa = 10^{-3}$ (top panel) and $\kappa = 10^{-3}i$ (bottom panel) and analytical fit with Eq. (10) (open circles), using $s_{ee}$, $s_{mm}$ from the left column.

These observations are in accord with the functional form of Eq.(10) and, to quantify them, we fit a Lorentzian function to $s_{ee}$, $s_{mm}$:

$$s_{ee/mm}(\omega) = \frac{ia_{e/m}\omega}{\omega_{e/m}^2 - \omega^2 + i\gamma_{e/m}\omega} + i\beta_{e/m}\omega \quad (14)$$

where the subscript $e/m$ denotes the respective parameters for the electric/magnetic conductivity (the modification of the Lorentzian is because the current is the time derivative of the dielectric polarization). In Fig. 3, left column, we plot $s_{ee}$ and $s_{mm}$ for $d = 130$ nm [repeated from Fig. 2(c), $\kappa = 0$] and their Lorentzian fits (open circles) with parameters $\omega_e = 2\pi \times 396.81$ THz, $\omega_m = 2\pi \times 394.22$ THz, $\gamma_e = 2\pi \times 1.85$ THz, $\gamma_m = 2\pi \times 2$ THz, $a_e = 2\pi \times 15.3$ THz, $a_m = 2\pi \times 27$ THz, $\beta_e = (2\pi)^{-1} \times 1.4$ fs and $\beta_m = 0$. The excellent fit of the Lorentzians to the electric and magnetic conductivities confirms the quasistatic nature of the electric and magnetic dipole resonances, respectively. In the right column of Fig. 3 we plot $\sigma_c$ for $\kappa = 10^{-3}$ and $\kappa = 10^{-3}i$ (top and bottom panel, respectively) and its analytical fit (open circles) with Eq.(10), i.e. $\sigma_c^{fit}(\omega) = \kappa C_0^{fit}(s_{ee} - i\beta_e\omega)(s_{mm} - i\beta_m\omega)$. In this expression $\kappa = 10^{-3}, 10^{-3}i$ is the chirality parameter used in the simulations, $C_0^{fit} = -0.268$ a constant which we use for the fitting and $\beta_e$, $\beta_m$ are the parameters used previously in the Lorentzians; $s_{ee}$, $s_{mm}$ are the retrieved conductivities (for $\kappa = 0$). The reason for subtracting the terms related to $\beta_e$, $\beta_m$ is that, besides the electric and magnetic mode, in the actual system there are other modes at nearby frequencies contributing a background, while in Eq.(10) $s_{ee}$, $s_{mm}$ are the resonant responses of exactly one electric and one magnetic mode, respectively [compare with the respective quantities given in Eq. (4)].

The excellent agreement between the numerically retrieved conductivity $\sigma_c$ and its analytical fit confirms the simple functional form of the magneto-electric coupling, as described by Eq.(10), and justifies the approximation $\sigma_c \ll s_{ee}, s_{mm}$, which is also evident in the relative values of $\sigma_c$ and $s_{ee}$, $s_{mm}$, shown in Fig. 3. We can further verify the approximation $D_e D_m + \kappa_c^2 a_e^{-1} a_m^{-1} \cong D_e D_m$, on which we were based in our coupled oscillator model to obtain the central result of Eq.(10). By combining Eq.(8) with Eq.(10) we find $\kappa_c^2 a_e^{-1} a_m^{-1} = -a_e a_m \omega^2 \kappa^2 C_0^2$, where $C_0 = C_0^{fit}/4$ [note that $s_{ee} s_{mm} = (\zeta \sigma_{ee}/2)(\sigma_{mm}/2\zeta) = \sigma_{ee}\sigma_{mm}/4$]. Then, using the fitting parameters $a_{e/m}$, $\gamma_{e/m}$, $\omega_{e/m}$ and $C_0^{fit}$, we find that the ratio $|(\kappa_c^2 a_e^{-1} a_m^{-1})/(D_e D_m)|$ is maximized around the resonant frequencies, $\omega_e$, $\omega_m$ to yield $\sim 10^{-7} \ll 1$, for both $\kappa = 10^{-3}, 10^{-3}i$. We note that the magnitude of $\kappa$ used in these examples is 2-3 orders stronger than that of natural chiral media, demonstrating that this approximation still holds even under such high values of $\kappa$.

### B. Retrieval of $\kappa$ in terms of the conductivities

Most notably, our model verifies the linearity of $\sigma_c$ in $\kappa$. This allows for the determination of an unknown chirality parameter of a chiral inclusion: we can use inclusion A of known chirality $\kappa_A$, as reference, to numerically retrieve $\sigma_{c,A}$, and inclusion B of unknown chirality $\kappa_B$ to retrieve $\sigma_{c,B}$. Then, because $\sigma_c \propto \kappa$, the unknown chirality will be simply given by $\kappa_B = \kappa_A (\sigma_{c,B}/\sigma_{c,A})$. This is illustrated in Fig. 4 for two inclusions A and B with dispersive chirality parameters $\kappa_A$ and $\kappa_B$, respectively, of the form [2,41]:

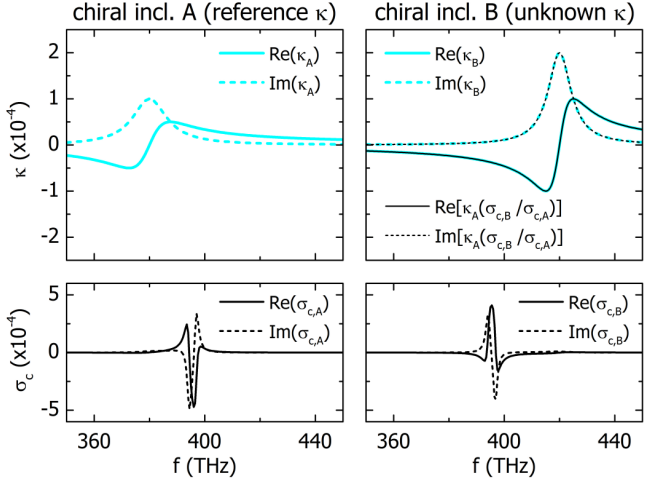

Figure 4: Retrieval of the unknown chirality parameter $\kappa_B$ of inclusion B, with reference chirality parameter $\kappa_A$ of inclusion A, using the sheet model conductivities. *Top row:* chirality parameters $\kappa_A$, $\kappa_B$ of the two inclusions (cyan lines) and retrieved $\kappa_B$ (black lines) of inclusion B. *Bottom row*: numerically retrieved conductivities $\sigma_c$ using Eq.(13c) from the sheet model, for the achiral isotropic metasurface of Fig. 2 with $d = 130$ nm.

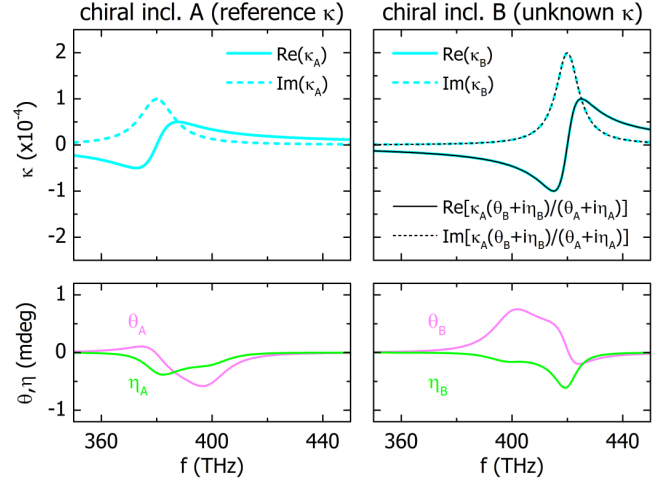

Figure 5: Retrieval of the unknown chirality parameter $\kappa_B$ of inclusion B, with reference chirality parameter $\kappa_A$ of inclusion A, using the chiroptical signals $\theta$, $\eta$ in transmission. *Top row:* chirality parameters $\kappa_A$, $\kappa_B$ of the two inclusions (cyan lines) and retrieved $\kappa_B$ (black lines) of inclusion B. *Bottom row*: numerically calculated chiroptical signals $\theta$, $\eta$ of the transmitted wave, for the achiral isotropic metasurface of Fig. 2 with $d = 130$ nm.

$$\kappa_{A/B}(\omega) = \frac{\omega_{\kappa,A/B}\omega}{\omega_{0\kappa,A/B}^2 - \omega^2 + i\gamma_{\kappa,A/B}\omega} \quad (15)$$

As an example, for inclusion A we have used $\omega_{0\kappa,A} = 2\pi\times380$ THz, $\gamma_{\kappa,A} = 2\pi\times15$ THz and $\omega_{\kappa,A} = 2\pi\times1.5\times10^{-3}$THz and, for inclusion B, $\omega_{0\kappa,B} = 2\pi\times420$ THz, $\gamma_{\kappa,B} = 2\pi\times10$ THz and $\omega_{\kappa,B} = 2\pi\times2\times10^{-3}$THz as shown in the top row panels of Fig. 4 (cyan lines). These parameters are chosen so that $\kappa_A$, $\kappa_B$ (a) are closer to realistic values of the chirality parameter of chiral molecules (e.g., aqueous solutions of monosaccharides [34,35] or biomolecules [14,23,32]) and (b) undergo changes both in sign and magnitude within the examined frequency range.

### C. Retrieval of $\kappa$ with measurements in transmission

In the previous procedure we used all reflection and transmission amplitudes to fully characterize the response of the metasurface with the chiral inclusions and consequently deduce the chirality parameter $\kappa$ of the latter. However, in most relevant experiments [24–32] chiral sensing relies on measurements in transmission only and, therefore, it would be useful to be able to retrieve the unknown $\kappa$ accordingly. With the transmission amplitudes given by Eqs. 12(a),(c) we can use the Stokes parameters to analyze the polarization state of the transmitted wave (see SI). After some calculations we find that the chiroptical signals of rotation $\theta$ and ellipticity $\eta$ of the transmitted wave, relative to the incident, are given by:

$$\theta = \frac{1}{2}\tan^{-1}\left(2\operatorname{Re}\left(\frac{\sigma_c}{s_{ee}s_{mm}-1}\right)\right) \cong \operatorname{Re}\left(\frac{\sigma_c}{s_{ee}s_{mm}-1}\right) \quad (16a)$$

$$\eta = \frac{1}{2}\tan^{-1}\left(2\operatorname{Im}\left(\frac{\sigma_c}{s_{ee}s_{mm}-1}\right)\right) \cong \operatorname{Im}\left(\frac{\sigma_c}{s_{ee}s_{mm}-1}\right) \quad (16b)$$

where we have used $\sigma_c \ll s_{ee}, s_{mm}$ to approximate $\tan^{-1}(x)\sim x$. We note here that even for signals as large as ~5 deg the error in this approximation is in the order of only 0.25% (in practice, the $\theta,\eta$ signals are in the order of some mdeg [31]). Taking into account the linearity of $\sigma_c$ in $\kappa$, as we found with our sheet model [see Eq.(10)], the results of Eqs.(16a),(16b) demonstrate that $\theta,\eta$ are proportional to $\kappa$ and, therefore, justify why *measurements of $\theta,\eta$ are sensitive to both $sgn(\kappa)$ and $|\kappa|$, and to both Re($\kappa$) and Im($\kappa$)*. In particular, according to Eqs.(16a) and (16b), $\theta$ and $\eta$ are the real and imaginary part, respectively, of the same quantity, $\sigma_c/(s_{ee}s_{mm}-1)$. Therefore, we may combine the two expressions and solve for $\sigma_c$, to obtain:

$$\sigma_c = (\theta + i\eta)(s_{ee}s_{mm} - 1) \quad (17)$$

The result of Eq. (17) is of great importance, as it enables us to unambiguously determine an unknown chirality solely from the chiroptical signals $\theta$, $\eta$ in transmission. To see how this is possible, let us assume the chiroptical signals $\theta_A,\eta_A$ and $\theta_B,\eta_B$, from simulations or experiments with inclusions A and B, respectively, which correspond to the conductivities $\sigma_{c,A}$ and $\sigma_{c,B}$, according to Eq.(17). Taking into account that $\sigma_c \propto \kappa$, we can divide the two conductivities to eliminate the common $s_{ee}s_{mm} - 1$ term and express the unknown $\kappa_B$ in terms of the reference $\kappa_A$, as:

$$\kappa_B = \kappa_A \frac{\theta_B + i\eta_B}{\theta_A + i\eta_A} \quad (18)$$

This is illustrated in Fig. 5 for the two inclusions A and B previously considered in Fig. 4. The chirality parameters of the two inclusions are shown in the top row panels of Fig. 5 (cyan lines) and the calculated chiroptical signals $\theta$, $\eta$ for disk height $d = 130$ nm are shown in the bottom row panels of Fig. 5 (magenta and green lines, respectively). The retrieved chirality of $\kappa_B$ using Eq.(18) is shown as black lines in the same panel with $\kappa_B$. The result of Eq.(18) is particularly practical for experiments and applications. In comparison to the previous approach, now an intermediate step involving the calculation of effective conductivities is not required and, importantly, only the transmitted fields are needed for the direct measurement of $\kappa$.

### D. Circular dichroism measurements

Alternatively, in view of circular dichroism (CD) measurements that are met both in traditional polarimetry [3] and contemporary nanophotonic approaches [24–30,32], we can relate our model with measurements of Right/Left Circularly Polarized (RCP/LCP or +/-) illumination. CD expresses the preferential absorption between ± waves and, in media for which the reflected power is the same for both ± waves, CD is characterized by $\Delta T$, the differential transmittance; with $T_\pm$, $R_\pm$ and $A_\pm$ denoting the transmitted, reflected and absorbed power for ± waves, respectively, one can easily see that $A_+ - A_- = (1-T_+-R_+)-(1-T_--R_-) = -T_++T_- = -\Delta T$. The transmission and reflection amplitudes of RCP/LCP waves can be derived directly from Eqs. (12a)-(12d), yielding (see SI for details):

$$t_{++} = \frac{1 - s_{ee}s_{mm} + i\sigma_c}{(1+s_{ee})(1+s_{mm})} \quad (19a)$$

$$t_{--} = \frac{1 - s_{ee}s_{mm} - i\sigma_c}{(1+s_{ee})(1+s_{mm})} \quad (19b)$$

$$t_{+-} = t_{-+} = 0 \quad (19c)$$

$$r_{+-} = r_{-+} = \frac{s_{mm} - s_{ee}}{(1+s_{ee})(1+s_{mm})} \quad (19d)$$

$$r_{++} = r_{--} = 0 \quad (19e)$$

where the subscripts in $t_{out,inc}$, $r_{out,inc}$ denote the output and incident polarization, respectively. The differential transmittance $\Delta T \equiv T_+ - T_-$, where $T_\pm \equiv |t_{\pm\pm}|^2$, is calculated from Eqs. (19a),(19b):

$$\Delta T = 4\frac{\text{Im}(\sigma_c^*(1 - s_{ee}s_{mm}))}{|(1+s_{ee})(1+s_{mm})|^2} \quad (20)$$

According to this result, $\Delta T$ is proportional to $\sigma_c$ and consequently to $\kappa$ [$\sigma_c \propto \kappa s_{ee}s_{mm}$, see Eq.(10)]; therefore measurements of $\Delta T$ are sensitive to both sgn($\kappa$) and |$\kappa$|. Additionally, the detection of both Re($\kappa$) and Im($\kappa$) is possible, as both choices lead to a complex $\sigma_c$ and therefore to a nonzero $\Delta T$. The sensitivity of $\Delta T$ to Im($\kappa$) is not a surprise, because the CD is known to be sensitive to Im($\kappa$) [4]. However, the sensitivity of $\Delta T$ measurements to Re($\kappa$) can be particularly counter-intuitive, as this means that even a lossless chiral inclusion (purely real $\kappa$) can lead to a difference in absorptance between left- and right-circularly polarized waves. This observation is in contrast to traditional CD measurements of pure chiral films, which are sensitive only to Im($\kappa$), and is also beyond the analytical expressions developed in the works of Refs. [16,24,25], in which the CD signal is predicted to depend on Im($\kappa$) only. On the contrary, and in support of our findings, in [32] García-Guirado et al. study a similar system of dielectric nanodisks and, in order to explain their CD signals, they take Re($\kappa$) into account explicitly: *"While the real and imaginary parts of κ are related to the optical rotary dispersion (ORD) and CD, respectively, the sensors facilitate the means of conversion in between these components, effectively allowing Re[κ] to yield a CD signal"*. Additionally, in [14] Abdulrahman et al. study an achiral plasmonic metasurface, in which they explain their CD signals in terms of Re($\kappa$) ($\kappa$ is denoted as $\xi_c$, instead): *"Importantly, since Im[$\xi_c(\omega)$] is relatively small at $\omega \sim \omega_{plasmon}$, the plasmon-induced CD comes mostly from the real part of the chiral parameter, Re[$\xi_c(\omega)$]."* Our findings are in accord with these two recent experimental works and explain why CD measurements should be interpreted with care with respect to the magnitude and sign of $\kappa$, particularly at frequency ranges far from the molecular resonances, where Re($\kappa$) is significantly stronger than Im($\kappa$).

### E. Enhancement mechanism of chiroptical signals

The result of Eq.(10) clearly shows that, with reduced detuning between the electric and magnetic mode, the magneto-electric coupling increases and, according to the form of Eqs.(16a),(16b) and (20), this implies that the $\theta$, $\eta$ and $\Delta T$ signals are expected to be enhanced. To examine the enhancement of the chiroptical signals in terms of the mode detuning, we scan the disk height $d$ and in Fig. 6(a) we plot the absolute value of the co- and cross- transmission amplitudes $t$ and $t_c$, respectively, for $\kappa$ as previously considered in Fig.2, i.e. $\kappa = 10^{-3}$ and $\kappa = 10^{-3}i$. Note that |$t$| remains practically unchanged when we switch $\kappa$ on and |$t_c$| is the same for both $\kappa = 10^{-3}$ and $\kappa = 10^{-3}i$; the latter is

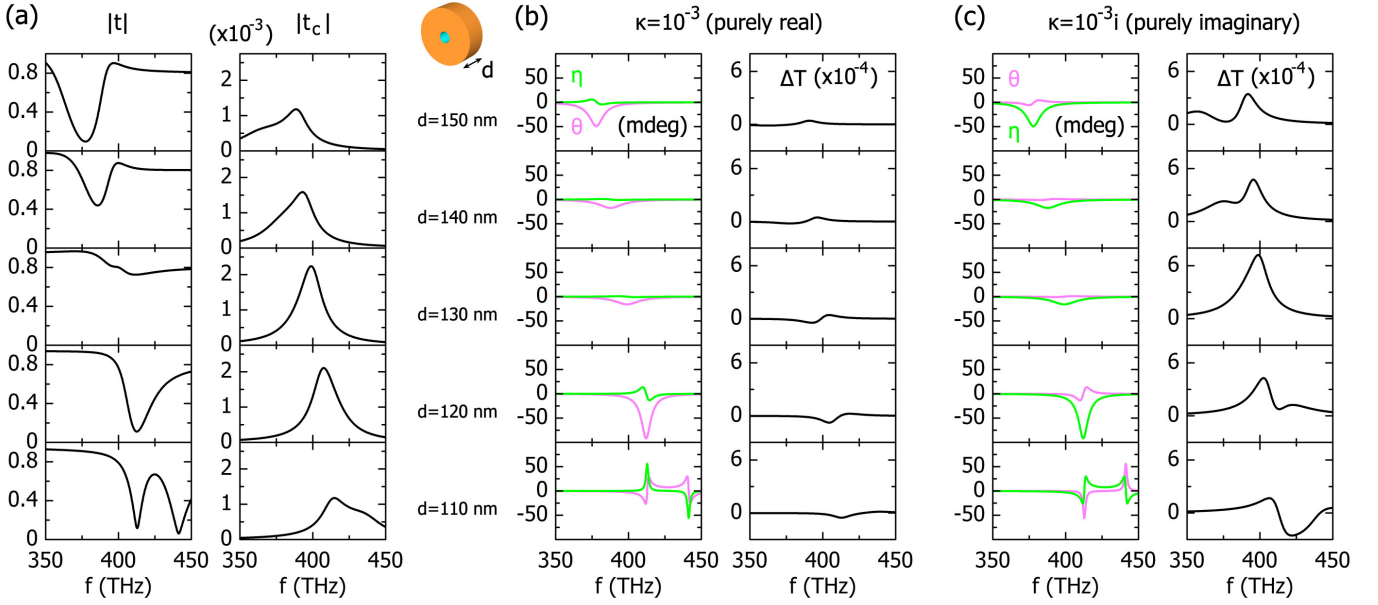

Figure 6: Response in transmission of achiral isotropic metasurface with chiral inclusions, for enhanced chiral sensing. (a) Absolute value of co- and cross- transmission amplitudes $t$ and $t_c$, respectively, for $\kappa = 10^{-3}$ or $\kappa = 10^{-3}i$. Chiroptical signals $\theta$ (magenta lines) and $\eta$ (green lines) and differential transmittance $\Delta T$ for (b) $\kappa = 10^{-3}$ and (c) $\kappa = 10^{-3}i$. The numerical results are in agreement with the analytical expressions.

expected from the form of Eq.(10), when combined with Eq.(12c). The chiroptical signals $\theta$, $\eta$ and $\Delta T$ for purely real and purely imaginary $\kappa$ are shown in Fig. 6(b) and Fig. 6(c), respectively.

First, we observe that as we scan the disk height $d$, for $d = 130$ nm where the magneto-electric coupling is maximized [see Fig. 2(d)], $\theta$ and $\eta$ become the weakest. This unexpected behavior can be explained with our theoretical model. From Eqs (12a),(12c) we can write $\sigma_c/(s_{ee}s_{mm}-1) = t_c/t$, and therefore express Eqs. (16a),(16b) as $\theta = \mathrm{Re}(t_c/t)$ and $\eta = \mathrm{Im}(t_c/t)$. According to this result, *the chiroptical signals $\theta$, $\eta$ change as $\sim t_c/t$*, that is, for a certain polarization conversion strength expressed by $t_c$, the far-field chiroptical signals are expected to increase inversely with transmittance $t$. Indeed, in the simulations we observe that, as the resonant frequencies of the modes become less detuned and the magneto-electric coupling (and therefore $|t_c|$) increases, the transmission amplitude $|t|$ increases exceedingly [see Fig. 6(a)]. In turn, this leads to reduced $\theta$, $\eta$, because, as predicted by our model, $\mathrm{Re}(t_c/t) \equiv \theta$ and $\mathrm{Im}(t_c/t) \equiv \eta$ (also verified numerically with calculating the ratio $t_c/t$ separately and taking its real and imaginary part). On the other hand, our model predicts that –unlike $\theta,\eta$– an increase in $t$ does not necessary have to lead to reduced $\Delta T$, due to the equivalent terms in the nominators and denominators of $t$ and $\Delta T$ [Eqs.(12a) and (20), respectively]. Indeed, in the simulations we observe that measurements of $\Delta T$ change roughly proportionally with $|t|$ for purely imaginary $\kappa$ [Fig. 6(c)], as also demonstrated in [25], while for real $\kappa$ the signal is significantly suppressed, regardless of $|t|$ [Fig. 6(b)]. For both real and imaginary $\kappa$, however, all signals are significantly stronger than those from the chiral inclusions alone. For example, in the absence of the nanodisks, chiral cylinders of radius 10 nm and height 130 nm (periodically arranged in air) give at 400 THz signals of $\theta = 0.06$ mdeg, $\eta = 0$ mdeg and $\Delta T = 0$ for $\kappa = 10^{-3}$ and $\theta = 0$ mdeg, $\eta = 0.06$ mdeg and $\Delta T = 4\times 10^{-6}$ for $\kappa = 10^{-3}i$.

Second, we observe that, each of $\mathrm{Re}(\kappa)$, $\mathrm{Im}(\kappa)$ manifests in both $\theta$, $\eta$ signals. The reason is that, because $\sigma_c \propto \kappa s_{ee} s_{mm}$ [see Eq.(10)], a purely real or a purely imaginary $\kappa$ results in a complex-valued $\sigma_c$ ($s_{ee}$, $s_{mm}$ are complex in general), therefore leading to both nonzero $\theta$ and $\eta$ [see Eq.(16)]. The same observation holds for $\Delta T$, and is also captured by Eq.(20), where a purely real or a purely imaginary $\kappa$ yields a complex-valued $\sigma_c$ and therefore nonzero $\Delta T$. These observations, which are predicted by our theoretical model, are in contrast to typical polarimetric measurements, where $\mathrm{Re}(\kappa)$ manifests only in $\theta$ ($\eta = 0$, $\Delta T = 0$) and $\mathrm{Im}(\kappa)$ only in $\eta$ and $\Delta T$ ($\theta=0$) [3]. In other words, the far-field $\theta$, $\eta$ and $\Delta T$ signals involve contributions from both the chiral inclusions and the metasurface and, therefore, should be interpreted with care in relevant experiments.

Last, for all examined disk heights we observe that $\theta(i\kappa) = -\eta(\kappa)$, $\eta(i\kappa) = +\theta(\kappa)$, which is a direct consequence of the fact that $\theta$, $\eta$ are the real and imaginary part, respectively, of the same quantity $\Phi = \sigma_c/(s_{ee}s_{mm}-1)$, as predicted by our sheet model: because $\sigma_c \propto \kappa s_{ee} s_{mm}$, the substitution $\kappa \to i\kappa$ leads to $\sigma_c \to i\sigma_c$ and therefore $\Phi \to i\Phi$; as $\mathrm{Re}(i\Phi) = -\mathrm{Im}(\Phi)$, $\mathrm{Im}(i\Phi) = \mathrm{Re}(\Phi)$ for any complex number $\Phi$, this leads to $\theta(i\kappa) = -\eta(\kappa)$, $\eta(i\kappa) = \theta(\kappa)$. This distinct behavior is what enables us to distinguish $\mathrm{Re}(\kappa)$ and $\mathrm{Im}(\kappa)$, as demonstrated in Fig. 5.

## IV. DISCUSSION

Our model [Eqs.(10),(16),(17),(20)], despite its simplicity, captures the underlying mechanism of chiral sensing with achiral metasurfaces: the far-field signals $\theta$, $\eta$ and $\Delta T$ are proportional to the magneto-electric conductivity $\sigma_c$, which in turn is proportional to the chirality parameter $\kappa$ of the chiral inclusions [Eq.(10)]. Therefore such measurements are sensitive to $|\kappa|$ and sgn($\kappa$) and to Re($\kappa$) and Im($\kappa$). While the far-field signals $\theta$, $\eta$ allow for the unambiguous determination of Re($\kappa$) or Im($\kappa$), measurements of differential transmittance, $\Delta T$, which are intuitively expected to be sensitive only to Im($\kappa$), are sensitive to Re($\kappa$), as well, as also recently shown experimentally in [14,32]. As we showed with Eq. (20), this is justified within the context of nanophotonic-based sensing schemes, as $\Delta T$ results from the combined response of the chiral inclusions and the metasurface. The fact that, in our example, the $\Delta T$ signals are dominated by Im($\kappa$) [compare $\Delta T$ between Fig.6(b) and Fig.6(c)] is captured by Eq.(20); close to the resonant frequency of the electric or the magnetic mode, the conductivity becomes purely real and, hence, when the detuning between the two modes vanishes, $\Delta T$ becomes simply the imaginary part of a practically real quantity multiplied by $\kappa$. Therefore, for purely real $\kappa$, $\Delta T$ is much weaker than for purely imaginary $\kappa$, thus justifying the big discrepancy between the two cases shown in Fig. 6. However, this does not have to be the case always, as the real or the imaginary part of $\sigma_c$ will be, in general, different for other configurations.

Besides providing insight to the sensitivity to the sign of $\kappa$ and to the individual contributions of Re($\kappa$) and Im($\kappa$), our model captures the mechanism of chiroptical signal enhancement: the far-field signals $\theta$, $\eta$ and $\Delta T$ are proportional to the magneto-electric conductivity $\sigma_c$, which in turn is proportional to the spatial overlap of the electric and magnetic mode, and to the product of their individual conductivities, $\sigma_{ee}\sigma_{mm}$ [Eq.(10)]. These are properties of the metasurface alone and can be, therefore, tuned independently from the chiral inclusion to enhance the weak chirality parameter $\kappa$.

In particular, the spatial overlap of the modes as expressed via the integral in Eq.(9) is directly related to the optical chirality density $C$ in the same region – one can easily show that $\mathrm{Im}(\mathbf{E}_m^*\mathbf{H}_e - \mathbf{E}_e\mathbf{H}_m^*) \equiv \mathrm{Im}(\mathbf{E}^*\mathbf{H}) \propto C$, where $\mathbf{E},\mathbf{H}$ are the total fields, that is, $\mathbf{E} = \mathbf{E}_e+\mathbf{E}_m$, $\mathbf{H} = \mathbf{H}_e+\mathbf{H}_m$ (see SI). With inspection of Eq.(9), this result associates the coupling constant $\kappa_c$ in our coupled oscillator model with $\kappa\times C$, i.e. the product of the chirality parameter $\kappa$ and the optical chirality density $C$. The importance of this observation is twofold. First, it associates our approach, according to which the two modes are coupled via $\kappa$, to the equivalent approach, according to which $\kappa$ couples to the combined excitation of both modes. In the latter case, $C$ expresses the coupling strength between the combined excitation of both modes and $\kappa$, while in the former case $C$ expresses the coupling strength between the two modes, which is mediated by $\kappa$. Second, it associates the enhancement of $C$ to the enhanced detection of both Re($\kappa$) and Im($\kappa$). This is an unexpected result, because, as was shown in the seminal work of Tang and Cohen [4], the rate of excitation of a chiral molecule is proportional to the product Im($\kappa$)×$C$, and $C$ is therefore expected to enhance the detection of Im($\kappa$) only. Our model does not contradict such a fundamental principle, rather simply emphasizes that the chiroptical signals originate from an *effective chirality*, which contains contributions from both the chiral inclusions and the metasurface [see Eq.(7)]. As we demonstrated, a purely real $\kappa$ yields a complex magneto-electric conductivity $\sigma_c$, i.e. a complex effective chirality, and consequently nonzero circular dichroism, which is expressed as nonzero differential transmittance $\Delta T$. Therefore, our analysis generalizes the recipe met in most contemporary approaches, which aim to enhance Im($\kappa$) via enhancement of $C$; enhancement of $C$ can lead to enhanced circular dichroism signals for both Re($\kappa$) and Im($\kappa$).

## V. CONCLUSION

In this work we examined theoretically how enhanced sensing of the total chirality parameter $\kappa$ is possible with achiral isotropic metasurfaces. By employing a simple thin-sheet model we derived analytical expressions, which we verified with numerical calculations of realistic systems. We showed that the far-field signals $\theta$, $\eta$ and $\Delta T$ of the composite system, i.e. metasurface with chiral inclusions, are proportional to the magneto-electric conductivity $\sigma_c$, which in turn is proportional to the chirality parameter $\kappa$ of the chiral inclusions. As a consequence, the far-field signals $\theta$, $\eta$ and $\Delta T$ are sensitive to both the magnitude and sign of $\kappa$ and to both Re($\kappa$) and Im($\kappa$). Regarding measurements of $\Delta T$, in particular, because these are intuitively expected to be sensitive to Im($\kappa$) only, they should be interpreted with care with respect to the magnitude and sign of $\kappa$, particularly at frequency ranges far from the molecular resonances, where Re($\kappa$) is significantly stronger than Im($\kappa$). We also showed that the magneto-electric conductivity $\sigma_c$ depends on the spatial and spectral properties of the electric and magnetic mode of the metasurface. As a consequence, the properties of the metasurface can be tailored independently from the chiral inclusion to enhance the far-fields signal, that is, the effective chirality of the composite system. Last, we proposed a scheme to unambiguously determine an unknown chirality in terms of a known (reference) chirality, using measurements of $\theta$ and $\eta$ solely in transmission. The generality of our approach extends beyond the specific examples, as our results apply to any homogenizable metasurface of subwavelength thickness that is combined with chiral inclusions of relatively weak chirality parameter $\kappa$.

## ACKNOWLEDGMENTS

This work was supported by the European Commission Horizon 2020, project ULTRACHIRAL (Grant No. FETOPEN-737071). The author is grateful to L. Bougas for discussions on experimental aspects of chiral sensing and to T. Koschny for discussions on theoretical aspects related to the sheet model. The author would also like to thank both colleagues for invaluable suggestions on the structure of the paper.

# Supporting Information for

# "Chiral sensing with achiral isotropic metasurfaces"

Sotiris Droulias

*Institute of Electronic Structure and Laser, FORTH, 71110 Heraklion, Crete, Greece*

E-mail: sdroulias@iesl.forth.gr

## Contents

## 1. Constitutive relations for sheet model

Let us assume that a metasurface is homogenizable and can be, therefore, replaced by a slab of thickness $D_{slab}$ and (bulk) effective susceptibilities $\chi_{ij}^{eff}$, where $i,j = \{e,m\}$. Inside the homogeneous slab, the constitutive relations are expressed as $\mathbf{D} = \varepsilon_0 \varepsilon_{eff} \mathbf{E} - i(\kappa_{eff}/c)\mathbf{H}$, $\mathbf{B} = \mu_0 \mu_{eff} \mathbf{H} + i(\kappa_{eff}/c)\mathbf{E}$, where $\varepsilon_{eff}$, $\mu_{eff}$ and $\kappa_{eff}$ are the effective permittivity, permeability and chirality of the metasurface [here and throughout this work, we follow the $\exp(+i\omega t)$ convention]. To find the connection between these bulk quantities and the surface susceptibilities considered in the coupled oscillator model, we start by writing the bulk effective parameters as a sum of background (non-resonant) and resonant contributions, that is, $\varepsilon_{eff} = \varepsilon_0(1+ \chi_{ee}^{eff})$, $\mu_{eff} = \mu_0(1+ \chi_{mm}^{eff})$ and $\kappa_{eff} = \kappa + \chi_{em}^{eff} = \kappa + \chi_{me}^{eff}$, where $\varepsilon_0$, $\mu_0$ are the vacuum permittivity and permeability, respectively, and $\kappa$ is the bulk chirality parameter of the chiral inclusions (the term '*resonant*' refers to the metasurface and not to the possible molecular resonances of the chiral inclusion). The constitutive equations are written accordingly, as:

$$\begin{pmatrix} \mathbf{D} \\ \mathbf{B} \end{pmatrix} = \overbrace{\begin{pmatrix} \varepsilon_0 & -i\dfrac{\kappa}{c} \\ +i\dfrac{\kappa}{c} & \mu_0 \end{pmatrix}}^{non-resonant} \begin{pmatrix} \mathbf{E} \\ \mathbf{H} \end{pmatrix} + \overbrace{\begin{pmatrix} \varepsilon_0 \chi_{ee}^{eff} & -i\dfrac{\chi_{em}^{eff}}{c} \\ +i\dfrac{\chi_{me}^{eff}}{c} & \mu_0 \chi_{mm}^{eff} \end{pmatrix}}^{resonant} \begin{pmatrix} \mathbf{E} \\ \mathbf{H} \end{pmatrix} \tag{S1}$$

The resonant susceptibilities $\chi_{ee}^{eff}, \chi_{mm}^{eff}, \chi_{em}^{eff}, \chi_{me}^{eff}$ are the bulk susceptibilities (dimensionless) due to the resonant modes. In particular $\chi_{em}^{eff} = \chi_{me}^{eff}$ is the contribution to the chirality parameter originating from the coupling of the resonant modes with the chiral inclusions; in the absence of coupling, $\chi_{em}^{eff} = \chi_{me}^{eff}$ vanishes, and the effective chirality $\kappa_{eff}$ reduces to $\kappa$. All $\chi_{ij}^{eff}$ are bulk quantities and are related to the surface susceptibilities $\chi_{ij}$ of our thin polarizable sheet model as $\chi_{ij}^{eff} = \chi_{ij} / D_{slab}$, i.e. $\chi_{ij}$ are measured in units of [m]. By expressing the resonant part of Eq.(S1) in terms of the surface susceptibilities, we can then relate them to the surface conductivities as:

$$\begin{pmatrix} \langle p_e \rangle \\ \langle p_m \rangle \end{pmatrix} = \begin{pmatrix} \varepsilon_0 \chi_{ee} & -i\dfrac{\chi_{em}}{c} \\ +i\dfrac{\chi_{me}}{c} & \mu_0 \chi_{mm} \end{pmatrix} \begin{pmatrix} E_{loc} \\ H_{loc} \end{pmatrix} \xleftarrow{\begin{smallmatrix} \langle j_{ee} \rangle = i\omega \langle p_{ee} \rangle = i\omega\varepsilon_0 \chi_{ee} E_{loc} \\ \langle j_{mm} \rangle = i\omega \langle p_{mm} \rangle = i\omega\mu_0 \chi_{mm} H_{loc} \\ \langle j_{em} \rangle = i\omega \langle p_{em} \rangle = i\omega(-i\chi_{em}/c) H_{loc} \\ \langle j_{me} \rangle = i\omega \langle p_{me} \rangle = i\omega(+i\chi_{me}/c) E_{loc} \end{smallmatrix}} \begin{pmatrix} \langle j_e \rangle \\ \langle j_m \rangle \end{pmatrix} = \begin{pmatrix} \sigma_{ee} & \sigma_{em} \\ \sigma_{me} & \sigma_{mm} \end{pmatrix} \begin{pmatrix} E_{loc} \\ H_{loc} \end{pmatrix} \quad (S2)$$

where $p$ denotes the dipole moments, with the subscript being in accord with the respective conductivities.

The conductivities $\sigma_{ee} = i\omega\varepsilon_0\chi_{ee}$, $\sigma_{mm} = i\omega\mu_0\chi_{mm}$ are measured in units of S and Ω, respectively and the magneto-electric conductivities $\sigma_{em} = i\omega(-i\chi_{em}/c) = (\omega/c)\chi_{em}$ and $\sigma_{me} = i\omega(+i\chi_{me}/c) = -(\omega/c)\chi_{me}$ are dimensionless.

## 2. Connection between the optical chirality density $C$ and the coupling constant $\kappa_c$

Let us denote with ($\mathbf{E}_e$, $\mathbf{H}_e$) and ($\mathbf{E}_m$, $\mathbf{H}_m$) the fields of the electric and magnetic mode, respectively, that drive the local surface currents $j_e$ and $j_m$. The total electric and magnetic field will be given by the sum $\mathbf{E} = \mathbf{E}_e + \mathbf{E}_m$ and $\mathbf{H} = \mathbf{H}_e + \mathbf{H}_m$, respectively, and the optical chirality density $C = -(\omega/2)\text{Im}(\mathbf{D}^*\mathbf{B})$ can be written in terms of the individual fields as:

$C \sim \text{Im}(\mathbf{E}^*\mathbf{H})$

$= \text{Im}[(\mathbf{E}_e + \mathbf{E}_m)^*(\mathbf{H}_e + \mathbf{H}_m)] =$

$= \text{Im}(\cancel{\mathbf{E}_e^*\mathbf{H}_e} + \mathbf{E}_e^*\mathbf{H}_m + \mathbf{E}_m^*\mathbf{H}_e + \cancel{\mathbf{E}_m^*\mathbf{H}_m}) =$ | $\mathbf{E}_e^*\mathbf{H}_e = 0$ and $\mathbf{E}_m^*\mathbf{H}_m = 0$ ($\mathbf{E}_e \perp \mathbf{H}_e$ and $\mathbf{E}_m \perp \mathbf{H}_m$)

$= \text{Im}(\mathbf{E}_e^*\mathbf{H}_m + \mathbf{E}_m^*\mathbf{H}_e) =$

$= \text{Im}(\mathbf{E}_e^*\mathbf{H}_m - \mathbf{E}_e\mathbf{H}_m^* + \mathbf{E}_e\mathbf{H}_m^* + \mathbf{E}_m^*\mathbf{H}_e) =$ | we add and subtract the term $\mathbf{E}_e\mathbf{H}_m^*$

$= \text{Im}(\mathbf{E}_m^*\mathbf{H}_e - \mathbf{E}_e\mathbf{H}_m^*) + \text{Im}(\mathbf{E}_e^*\mathbf{H}_m + \mathbf{E}_e\mathbf{H}_m^*) =$

$= \text{Im}(\mathbf{E}_m^*\mathbf{H}_e - \mathbf{E}_e\mathbf{H}_m^*) + \cancel{\text{Im}[2\text{Re}(\mathbf{E}_e^*\mathbf{H}_m)]} =$ | $\text{Im}[2\text{Re}(\mathbf{E}_e^*\mathbf{H}_m)] = 0$ [$\text{Re}(\mathbf{E}_e^*\mathbf{H}_m)$ is a purely real quantity]

$= \text{Im}(\mathbf{E}_m^*\mathbf{H}_e - \mathbf{E}_e\mathbf{H}_m^*)$

Hence $\text{Im}(\mathbf{E}_m^*\mathbf{H}_e - \mathbf{E}_e\mathbf{H}_m^*) \equiv \text{Im}(\mathbf{E}^*\mathbf{H}) \propto C$, and $C$ is therefore associated with the coupling constant $\kappa_c$ of our oscillator model.

## 3. Solution of Maxwell's equations for a 1D current sheet

To find the analytical expressions for the transmission amplitudes we start with replacing our metasurface by a thin polarizable sheet that extends on the *xy*-plane and supports electric and magnetic currents. We assume a wave propagating in region (1) along the +z direction that arrives at the sheet and is partly reflected and partly transmitted to region (2), as shown in Fig. A1.

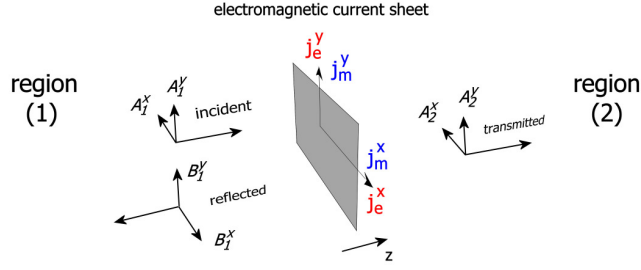

Figure A1: Schematic of a 1D electromagnetic sheet that supports electric ($j_e$) and magnetic ($j_m$) currents with components along the $x$, $y$ axes. The sheet separates the surrounding homogeneous space into regions (1) and (2) with different, in general, properties. The incident field (components $A_1^x, A_1^y$) propagates along the $+z$ direction and is reflected (components $B_1^x, B_1^y$) and transmitted (components $A_2^x, A_2^y$).

In the general case, the half-space on either side of the sheet has different material properties. For z<0 it is characterized by permittivity and permeability $\varepsilon_1$, $\mu_1$, respectively, and $\varepsilon_2$, $\mu_2$ for z>0. The current sheet is characterized by the conductivity tensor as given by Eq.(11), which relates the electric and magnetic current densities with the local fields as:

$$\begin{pmatrix} \langle j_e^x \rangle \\ \langle j_e^y \rangle \\ \langle j_m^x \rangle \\ \langle j_m^y \rangle \end{pmatrix} = \begin{pmatrix} \sigma_{ee} & 0 & \sigma_c & 0 \\ 0 & \sigma_{ee} & 0 & \sigma_c \\ -\sigma_c & 0 & \sigma_{mm} & 0 \\ 0 & -\sigma_c & 0 & \sigma_{mm} \end{pmatrix} \begin{pmatrix} E_{loc}^x \\ E_{loc}^y \\ H_{loc}^x \\ H_{loc}^y \end{pmatrix} \quad (S3)$$

Since the current sheet carries electric as well as magnetic currents, both the electric and magnetic fields have discontinuities at the current sheet. Therefore, the local fields must be defined as an average across the current sheet [1,2]:

$$\begin{pmatrix} E_{loc}^x \\ E_{loc}^y \\ H_{loc}^x \\ H_{loc}^y \end{pmatrix} = \frac{1}{2} \begin{pmatrix} E_{loc,1}^x + E_{loc,2}^x \\ E_{loc,1}^y + E_{loc,2}^y \\ H_{loc,1}^x + H_{loc,2}^x \\ H_{loc,1}^y + H_{loc,2}^y \end{pmatrix} = \frac{1}{2} \begin{pmatrix} A_1^x + B_1^x + A_2^x \\ A_1^y + B_1^y + A_2^y \\ -\frac{1}{\zeta_1} A_1^y + \frac{1}{\zeta_1} B_1^y - \frac{1}{\zeta_2} A_2^y \\ +\frac{1}{\zeta_1} A_1^x - \frac{1}{\zeta_1} B_1^x + \frac{1}{\zeta_2} A_2^x \end{pmatrix} \quad (S4)$$

where $\zeta_{1,2} = \sqrt{\frac{\mu_0 \mu_{1,2}}{\varepsilon_0 \varepsilon_{1,2}}} = \sqrt{\frac{\mu_0}{\varepsilon_0}} \sqrt{\frac{\mu_{1,2}}{\varepsilon_{1,2}}} = \zeta_0 \sqrt{\frac{\mu_{1,2}}{\varepsilon_{1,2}}}$ is the wave of each of the two half-spaces. The equivalent boundary conditions for the current sheet are $\mathbf{n} \times (\mathbf{E}_2 - \mathbf{E}_1) = -\langle \mathbf{j}_m \rangle$, $\mathbf{n} \times (\mathbf{H}_2 - \mathbf{H}_1) = +\langle \mathbf{j}_e \rangle$, where $\mathbf{n}$ is the surface normal of the current sheet pointing from region 1 to region 2. Application of the boundary conditions at $z = 0$, where the sheet is located, yields the system of equations:

$$\begin{pmatrix} \frac{1}{\zeta_1} & -\frac{1}{\zeta_1} & -\frac{1}{\zeta_2} & \frac{1}{\zeta_2} & 0 & 0 & 0 & 0 \\ 0 & 0 & 0 & 0 & \frac{1}{\zeta_1} & -\frac{1}{\zeta_1} & -\frac{1}{\zeta_2} & \frac{1}{\zeta_2} \\ 0 & 0 & 0 & 0 & -1 & -1 & 1 & 1 \\ 1 & 1 & -1 & -1 & 0 & 0 & 0 & 0 \end{pmatrix} \begin{pmatrix} A_1^x \\ B_1^x \\ A_2^x \\ 0 \\ A_1^y \\ B_1^y \\ A_2^y \\ 0 \end{pmatrix} = \begin{pmatrix} \langle j_e^x \rangle \\ \langle j_e^y \rangle \\ \langle j_m^x \rangle \\ \langle j_m^y \rangle \end{pmatrix} \quad (S5)$$

To cast the solution of Eq.(S5) in a compact form we define the wave impedance ratio $\rho \triangleq \frac{\zeta_2}{\zeta_1} = \frac{\sqrt{\mu_2/\varepsilon_2}}{\sqrt{\mu_1/\varepsilon_1}}$, ($\rho=1$ for the same material in both subspaces) and the normalized conductivities $s_{ee} = \zeta_2 \sigma_{ee}/2$, $s_{mm} = \sigma_{mm}/2\zeta_1$. Combining (S3) and (S4) into (S5) and solving for $A_2^x = t_{xx} A_1^x + t_{xy} A_1^y$, $A_2^y = t_{yy} A_1^y + t_{yx} A_1^x$, $B_1^x = r_{xx} A_1^x + r_{xy} A_1^y$ and $B_1^y = r_{yy} A_1^y + r_{yx} A_1^x$, we find that the scattering amplitudes $t_{xx}$, $t_{yy}$, $t_{xy}$, $t_{yx}$, $r_{xx}$, $r_{yy}$, $r_{xy}$, $r_{yx}$ are given by:

$$t \equiv t_{xx} = t_{yy} = \frac{\rho - s_{ee} s_{mm}}{\frac{\rho+1}{2} + s_{ee} + s_{mm} + \frac{\rho+1}{2\rho} s_{ee} s_{mm}} \tag{S6}$$

$$t_c \equiv t_{xy} = -t_{yx} = \frac{\rho \sigma_c}{\frac{\rho+1}{2} + s_{ee} + s_{mm} + \frac{\rho+1}{2\rho} s_{ee} s_{mm}} \tag{S7}$$

$$r \equiv r_{xx} = r_{yy} = \frac{s_{mm} - s_{ee} + \frac{\rho-1}{2} + \frac{\rho-1}{2\rho} s_{ee} s_{mm}}{\frac{\rho+1}{2} + s_{ee} + s_{mm} + \frac{\rho+1}{2\rho} s_{ee} s_{mm}} \tag{S8}$$

$$r_c \equiv r_{xy} = r_{yx} = 0 \tag{S9}$$

To obtain these results we have used the approximation $\sigma_c \ll \sigma_{ee}, \sigma_{mm}$, thus eliminating all terms of the form $\sigma_c^2$. Solving Eqs. (S6)-(S9) in terms of the conductivities, we obtain:

$$s_{ee} = \frac{\zeta \sigma_{ee}}{2} = \frac{\rho - \rho r - t}{1 + r + t} \tag{S10}$$

$$s_{mm} = \frac{\sigma_{mm}}{2\zeta} = \frac{\rho + \rho r - \rho t}{\rho - \rho r + t} \tag{S11}$$

$$\sigma_c = \frac{2 t_c (1 + \rho + r - \rho r)}{(\rho - \rho r + t)(1 + r + t)} \tag{S12}$$

For $\zeta_1 = \zeta_2$, i.e. $\rho = 1$, Eqs. (S6)-(S9) and Eqs.(S10)-(S12) become Eqs.(12a)-(12d) and Eqs.(13a)-(13c), respectively, presented in the main text.

## 4. Differential transmittance measurements

To find the differential transmittance in terms of circlularly polarized waves we can convert the linear polarization amplitudes to circular, using the following equations [3]:

$$\begin{pmatrix} t_{++} & t_{+-} \\ t_{-+} & t_{--} \end{pmatrix} = \frac{1}{2} \begin{pmatrix} (t_{xx} + t_{yy}) + i(t_{xy} - t_{yx}) & (t_{xx} - t_{yy}) - i(t_{xy} + t_{yx}) \\ (t_{xx} - t_{yy}) + i(t_{xy} + t_{yx}) & (t_{xx} + t_{yy}) - i(t_{xy} - t_{yx}) \end{pmatrix} \tag{S13}$$

$$\begin{pmatrix} r_{++} & r_{+-} \\ r_{-+} & r_{--} \end{pmatrix} = \frac{1}{2} \begin{pmatrix} (r_{xx} - r_{yy}) + i(r_{xy} + r_{yx}) & (r_{xx} + r_{yy}) - i(r_{xy} - r_{yx}) \\ (r_{xx} + r_{yy}) + i(r_{xy} - r_{yx}) & (r_{xx} - r_{yy}) - i(r_{xy} + r_{yx}) \end{pmatrix} \tag{S14}$$

The circular transmission amplitudes, $t_{++}$, $t_{-+}$, $t_{+-}$ and $t_{--}$, where the first subscript indicates the transmitted field polarization (±, RCP/LCP) and the second subscript indicates the incident field polarization, are then given by:

$$t_{++} = \frac{\rho - s_{ee}s_{mm} + i\rho\sigma_c}{\frac{\rho+1}{2} + s_{ee} + s_{mm} + \frac{\rho+1}{2\rho}s_{ee}s_{mm}} \tag{S15}$$

$$t_{--} = \frac{\rho - s_{ee}s_{mm} - i\rho\sigma_c}{\frac{\rho+1}{2} + s_{ee} + s_{mm} + \frac{\rho+1}{2\rho}s_{ee}s_{mm}} \tag{S16}$$

$$t_{+-} = t_{-+} = 0 \tag{S17}$$

Using Eq.(S15) and Eq.(S16) to calculate $\Delta T = |t_{++}|^2 - |t_{--}|^2$, we find:

$$\Delta T = 4\rho \frac{\mathrm{Im}\left(\sigma_c^*(\rho - s_{ee}s_{mm})\right)}{\left|\frac{\rho+1}{2} + s_{ee} + s_{mm} + \frac{\rho+1}{2\rho}s_{ee}s_{mm}\right|^2} \tag{S18}$$

For $\rho=1$ we retrieve the result of Eq.(20).

Similarly for the circular reflection amplitudes, $r_{++}$, $r_{-+}$, $r_{+-}$ and $r_{--}$, we have:

$$r_{+-} = r_{-+} = \frac{s_{mm} - s_{ee} + \frac{\rho-1}{2} + \frac{\rho-1}{2\rho}s_{ee}s_{mm}}{\frac{\rho+1}{2} + s_{ee} + s_{mm} + \frac{\rho+1}{2\rho}s_{ee}s_{mm}} \tag{S19}$$

$$r_{++} = r_{--} = 0 \tag{S20}$$

From Eqs.(S19),(S20) it is evident that the reflected power is the same for both ± waves, justifying why measurements of $\Delta T$ are equivalent to CD measurements, as analyzed in the main text.

## 5. Stokes Parameters and chiroptical signals $\theta$, $\eta$

To characterize the polarization state of the transmitted wave, we analyze it in $x$, $y$ components and then calculate the Stokes parameters, which are defined as:

$$\begin{aligned} S_0 &= E_x E_x^* + E_y E_y^* \\ S_1 &= E_x E_x^* - E_y E_y^* \\ S_2 &= E_x E_y^* + E_y E_x^* \\ S_3 &= i\left(E_x E_y^* - E_y E_x^*\right) \end{aligned} \tag{S21}$$

With the Stokes parameters known we can obtain information about the polarization state of the transmitted wave, in terms of its optical rotation $\theta$ and its ellipticity $\eta$, using the following formulas:

$$\begin{aligned} \theta &= \frac{1}{2}\tan^{-1}\left(\frac{S_2}{S_1}\right) \\ \eta &= \frac{1}{2}\tan^{-1}\left(\frac{S_3}{\sqrt{S_1^2 + S_2^2}}\right) \end{aligned} \tag{S22}$$

# 6. Achiral isotropic metasurface of dielectric nanodisks: additional examples

Here we demonstrate the generality of Eq.(10), which expresses the magneto-electric coupling in terms of the individual conductivities of the isotropic metasurface of nanodisks. In Fig. A2 we plot the numerically retrieved conductivities $\sigma_c$ for $\kappa = 10^{-3}$ together with plots of $\sigma_c^{fit}(\omega) = \kappa C_0^{fit}(s_{ee} - i\beta_e\omega)(s_{mm} - i\beta_m\omega)$. In this expression $\kappa = 10^{-3}$ is the chirality parameter used in the simulations, $C_0^{fit} = -0.268$ a constant which we use for the fitting and $\beta_e = (2\pi)^{-1} \times 1.4$ fs and $\beta_m = 0$; $s_{ee}$ and $s_{mm}$ are the numerically retrieved conductivities as shown previously in Fig. 2, which we repeat here for convenience.

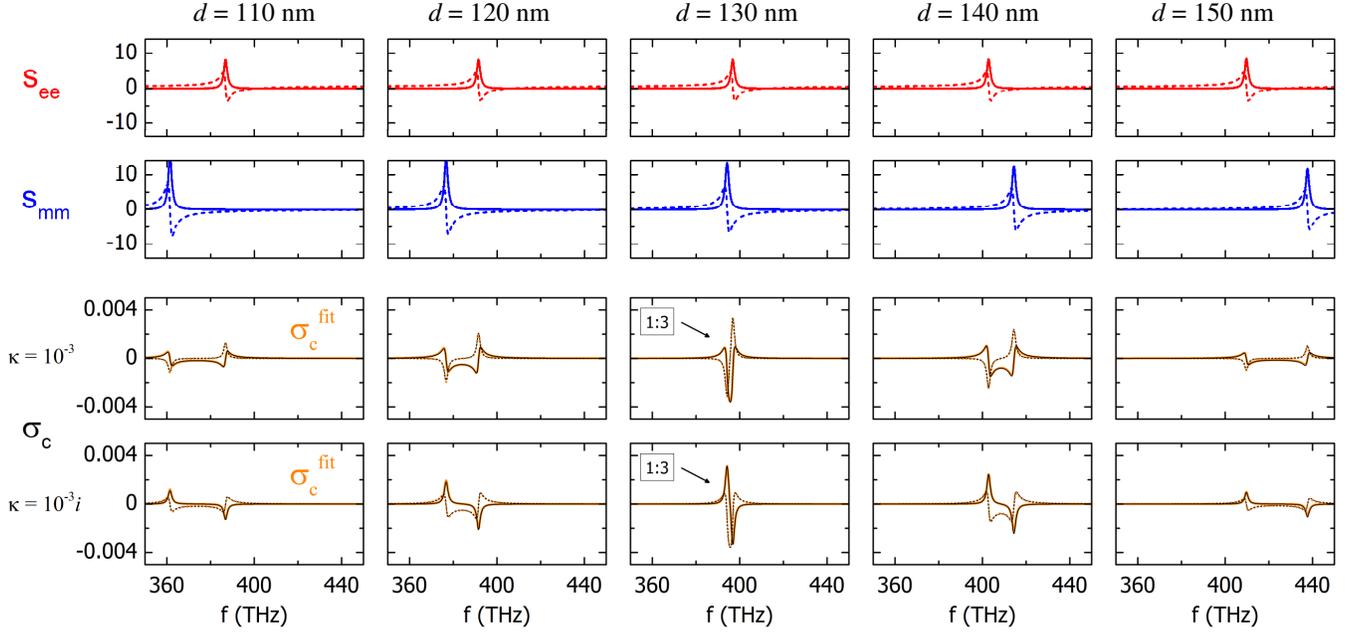

Figure A2: Numerically retrieved conductivities $s_{ee}$, $s_{mm}$ as a function of the disk height, $d$, for $\kappa = 0$ (top two rows) and numerically retrieved conductivity $\sigma_c$ for $\kappa = 10^{-3}$ (third row) and $\kappa = 10^{-3}i$ (fourth row) with analytical fit of $\sigma_c^{fit}$.

We emphasize that we determine the constants $C_0^{fit}$, $\beta_e$ and $\beta_m$ once and we do not change them afterwards as we repeat the calculations for different disk heights $d$. In other words, to calculate $\sigma_c^{fit}$ we only replace the numerically retrieved conductivities $s_{ee}$ and $s_{mm}$ for each $d$. The agreement between the numerically retrieved $\sigma_c$ (black lines) and the theoretically derived $\sigma_c^{fit}$ (orange lines) is excellent, validating the generality of Eq.(10) and the associated conclusions.

In Fig. A3 we repeat the plots of $\sigma_c$ and $\sigma_c^{fit}$ for $d = 130$ nm (top row), demonstrating all four possible cases for $\kappa$ purely real or imaginary, upon sign change. Again, we emphasize that once we fix the constants $C_0^{fit}$, $\beta_e$ and $\beta_m$ we only change $\kappa$, according to its value used in the simulations. In Fig. A3, top row, we plot both the numerically retrieved $\sigma_c$ and its analytical fit $\sigma_c^{fit}$ for each case. In the remaining panels of Fig. A3 we plot the chiroptical signals $\theta$, $\eta$ and the differential transmittance $\Delta T$ for the four examined cases. The solid and dashed lines are the numerically retrieved results and the open circles are the theoretically calculated quantities, using the retrieved conductivities, which we repeat here for convenience:

$$\Phi = \frac{\sigma_c}{s_{ee}s_{mm} - 1} \tag{S23}$$

$$\Delta T = 4\frac{\mathrm{Im}\left(\sigma_c^*(1 - s_{ee}s_{mm})\right)}{\left|(1 + s_{ee})(1 + s_{mm})\right|^2} \tag{S24}$$

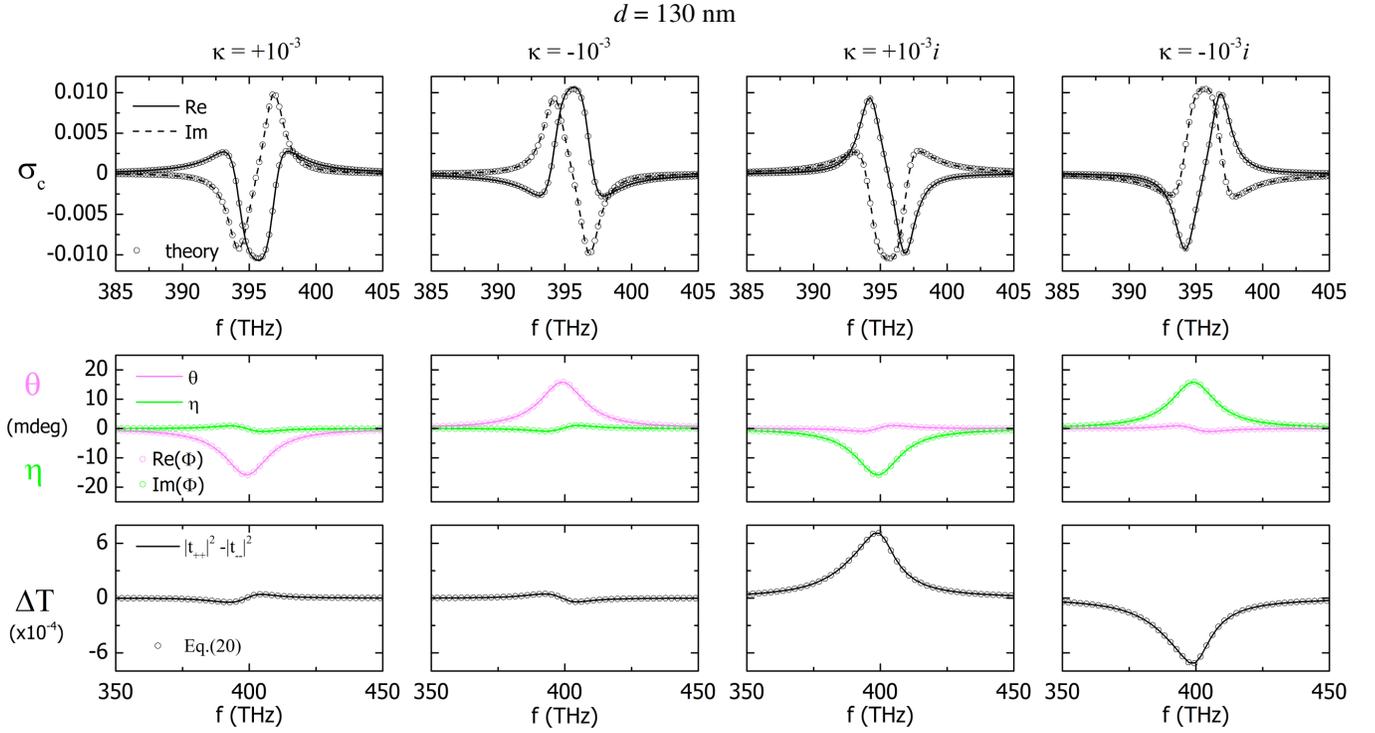

Figure A3: Numerically retrieved conductivity $\sigma_c$ and analytical fit of $\sigma_c^{fit}$ (top row) for $\kappa = \pm 10^{-3}$ (first, second column) and $\kappa = \pm 10^{-3}i$ (third, fourth column). The numerically retrieved chiroptical signals $\theta$, $\eta$ (middle row) and differential transmittance $\Delta T$ (bottom row) are shown as solid and dashed lines. The open circles are the theoretically calculated quantities, using the retrieved conductivities.